\documentclass[apj]{emulateapj}

\newcommand{\kpc}{{\rm kpc}}

\newcommand{\hmpc}{\ifmmode{h^{-1}\,\hbox{Mpc}}\else{$h^{-1}$\thinspace Mpc}\fi}

\newcommand{\kms}{\ifmmode{\rm \,\hbox{km}\,s^{-1}}\else {\rm\,km\,s$^{-1}$}\fi}
\newcommand{\msun}{{\rm\,M_\odot}}

\begin{document}
\title{Star Stream Folding by Dark Galactic Sub-Halos}
\shorttitle{Star Stream Folding}
\shortauthors{Carlberg}
\author{R.~G.~Carlberg}
\affil{Department of Astronomy and Astrophysics, \break 
University of Toronto, Toronto, ON M5S 3H4, Canada}
\email{carlberg@astro.utoronto.ca}

\begin{abstract}
Star streams in galactic halos are long, thin, unbound structures that will be disturbed by the thousands dark matter sub-halos that are predicted to be orbiting within the main halo. A sub-halo generally induces a localized wave in the stream which often evolves into a ``z-fold" as an initially trailing innermost part rotates faster than an initially leading outermost part. The folding, which becomes increasingly complex with time, leads to an apparent velocity dispersion increase and thickening of the stream. We measure the equivalent velocity dispersion around the local mean in the simulations, finding that it rises to about 10 \kms\ after 5 Gyr and 20 \kms\ after 13 Gyr. The currently available measurements of the velocity dispersion of halo star streams range from as small as 2 \kms\ to slightly over 20 \kms. The streams with velocity dispersions of 15-20 \kms\ are compatible with what sub-halo heating would produce. A dynamical understanding of the low velocity dispersion streams depends on the time since the progenitor's tidal disruption into a thin stream. If the streams are nearly as old as their stars then sub-halos cannot be present with the predicted numbers and masses. However, the dynamical age of the streams can be significantly less than the stars. If the three lowest velocity streams are assigned ages of 3 Gyr, they are in conflict with the sub-halo heating. The main conclusion is that star stream heating is a powerful and simple test for sub-halo structure.
\end{abstract}

\keywords{galaxies: dwarf - Local Group - dark matter}

\section{INTRODUCTION}
\nobreak
N-body simulations of individual galactic halos formed in a standard LCDM cosmology predict thousands of sub-halos orbiting within the main dark halo. Observations find that the numbers of dwarf galaxies per galaxy are at least a factor of ten less numerous than the sub-halos at masses of $~10^7\msun$ \citep{Mateo:98, Moore:99, Klypin:99,VL1, Aquarius,SG:07, Strigari:07, Carlberg:09}. Astrophysical solutions for the missing satellite problem accept that the predicted number of sub-halos are present and propose mechanisms which remove the gas early enough that the majority of the sub-halos contain no visible stars or gas. A leading idea notes that low mass dark halos are beginning to collapse around the time of reionization, so that only a fraction are able to initiate star formation before reionization heats up the gas \citep{Bullock:01, KGK, Munoz:09}. The more drastic alternative solution is to accept the discrepancy from the simulation predications as real, implying that the LCDM spectrum of density perturbations is suppressed on small scales so that fewer low mass dark halos form. 

The predicted multitude of orbiting dark sub-halos can, in principle, be detected through their gravitational effects. Gravitational lensing from sub-halos will lead to strong lensing and multiple images of background objects, although for typical redshifts the splitting occurs on scales of one-hundredth of an arc-second \citep{WP:92,Riehm:09} which is so small that its observation will require the next generation of telescopes. Sub-halos will disturb the disk of a galaxy \citep{Font:01, Ardi:03, Benson:04, Kazantzidis:08, Kazantzidis:09, Purcell:09} with the most recent simulations concluding that disk heating and morphological disturbances are substantial, but overlaid on the effects of dissipation in the gas and the internal dynamics,  so that it is not clear that sub-halos leave a unique signature. Sub-halos that orbit through the disk are themselves shock-heated and potentially disrupted \citep{DOSHK:09}.

The halos of galaxies contain thin, low velocity dispersion, star streams which are most likely the remnants of tidally disrupted infalling dwarf galaxies and globular clusters \citep{Quinn, Law:05}. \citet{JSH:02} recognized that the narrow spatial extent and low velocity dispersion of these streams made them sensitive indicators for the presence of dark sub-halos in the halo. They concluded on the basis of their analysis approach that the streams were not sufficiently disturbed to be able to constrain CDM models. \citet{SGV:08} noted that sub-halos induce clumps in tidal streams which potentially offered an observational test. Another look at the gravitational disturbances of streams from dark sub-halos is justified on the basis that recent simulations, with increased resolution, have found far more sub-halos (and a whole hierarchy of sub-sub-halos) than older simulations and provide more detailed overall information. Furthermore, ongoing observational work is discovering many new star streams both in the Milky Way and external galaxies. The streams are now identified with sufficient confidence that spectroscopic measurements of the internal velocity dispersion are being undertaken. This paper measures the sub-halo induced velocity dispersion increase, effectively a heating of the streams,  and compare the results to observations to assess whether the predicted numbers of sub-halos are compatible with star stream kinematics.

\section{SIMULATIONS}
\nobreak
We employ a somewhat idealized simulation, with a set of Monte Carlo realizations of the sub-halos, to allow a fairly broad exploration of general outcomes. Although the model halo is scaled to the Milky Way halo, our simulation does not include the visible galaxy components and is not intended to be a realistic match to the Milky Way. The model is useful to isolate and quantify the effects of sub-halos on streams, upon which the tidal disruption process and the more complex orbits within a real galaxy will add additional complexity. 

There are three dynamical components in our simulation. The fixed galaxy halo potential and a set of sub-halos together create the gravitational field in which a set of idealized star streams orbit. The sub-halos orbit as test particles and do not evolve with time, which in this case is a very conservative assumption, leaving out the vigour of the accretion history. The galaxy halo is modeled after the Aquarius simulation \citep{Aquarius} as a \citet{NFW} fixed spherical potential with a peak circular velocity of $v_{max}=210$ \kms\ at $r_{max}=30$ \kpc. A set of sub-halos are drawn from the $n(M)\propto M^{-1.9}$ distribution, with the maximum mass limited to $2\times 10^{10}\msun$ to limit large run to run variations. Initially the minimum sub-halo mass was arbitrarily set at $10^7\msun$, which created about 2513 satellite sub-halos. The total mass in these sub-halos is about 10\% of the mass of the main halo. We run 21 realizations of this model for our heating statistics.

Each sub-halo is modelled as a \citet{Hernquist} potential which is somewhat different than the best fit model to the Aquarius result but much simpler to use in our simulation. The scale radius relation is from the \citet{Aquarius} simulations, approximated as $a(M)= 6 (M/10^{10} \msun)^{0.43}\,\kpc$ \citep{Neto:07}. To test for sensitivity to the sub-halo structure we also use a constant concentration model approximately fit to the most massive sub-halos, $a(M)= 10.1 (M/10^{10} \msun)^{1/3}\,\kpc$, finding that the velocity dispersion is reduced by about a factor of 25\% at any time. The sub-sub-halos found in the simulations are not included.  This approach also ignores the gravitational wakes that massive objects induce in the background, although the dynamics of the disturbances in the rings suggests that the wakes are not likely to be very important for stream folding. The initial orbital velocities of the sub-halos are drawn from a non-rotating isotropic velocity dispersion distribution that satisfies the Jeans equation for the overall halo potential.

\begin{figure}
\epsscale{0.73}
\plotone{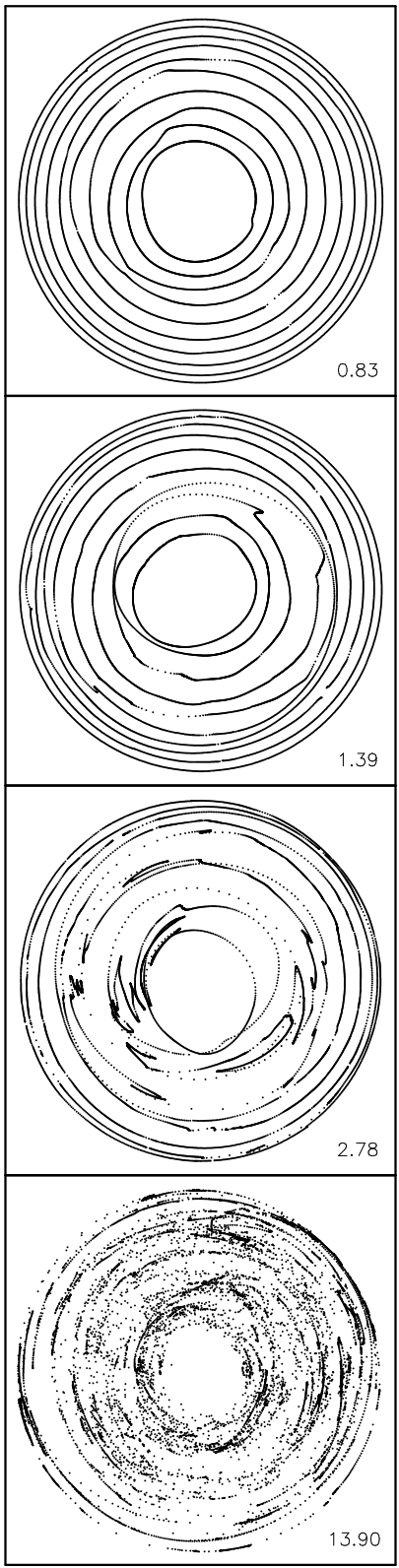}

\caption{Face on views of a time sequence within one simulation.  A range of stream locations are sampled with ten rings with 1000 particles per ring. These idealized streams are initiated on purely circular orbits which would preserve their appearance in the absence of sub-halos.
The time is shown in Gyr indicated in the lower right of each sub-panel. The plotted radii are the arctangent of the $r_{max}$ scaled radii to allow a more uniform display. Note how a wave-like perturbation evolves into a z-fold which gradually leads to a scrambling of the ring.
\label{fig_xy}}
\end{figure}

The star streams are idealized as a set of ten co-planar rings, with particles initially on circular orbits, that is, each ring starts with zero width and a purely tangential velocity equal to the circular velocity at that radius, see Fig.~\ref{fig_xy}. In the absence of sub-halos these circular rings do not change their appearance at all. The n-th ring is located at $0.5r_{max}(1.316)^{n-1}$, so the rings initially extend from 15 \kpc\ to 178 \kpc. Each ring is composed of 1000 equally spaced particles to provide adequate angular resolution as the streams are stretched.  
The rings are simple non-evolving distributions in the absence of sub-halos and cover the region where star streams are generally found, allowing us to investigate the dynamical effects from which we can infer the effects for a more general range of star streams. Future work will include a range of internal velocities and more general orbits.

The orbits of both the sub-halos and ring particles are integrated with a leap-frog scheme with a time step of $1.39\times 10^5$ year. At the peak circular velocity a particle will move 29 parsec in a time step, whereas the least massive of the 100 heaviest sub-halos has a scale radius of about 1.5 kpc, inside of which the acceleration has a constant magnitude, insuring that all motions are accurately computed. Even for sub-halos of $10^7\msun$ there are 10 steps per scale radius. We have repeated a test simulation with 5 times smaller steps and have confirmed that the orbits are identical to the precision of our analysis.  We have also used fewer sub-halos and conclude that the 2513 sub-halos reliably captures the measurable effects. A single run is 100,000 steps, equivalent to 13.9 Gyr, about a Hubble time. In Figure~\ref{fig_xy} we show the face-on time development of the distribution of the particles in one example. This particular realization has substantially disrupted inner rings. All of the 21 simulations show the characteristic ``z" folding of the rings which eventually leads to scrambled rings. 

\section{ANALYSIS}

The simplest analysis of the effects of sub-halos on our test particle rings is simply to measure the increase in radial velocity dispersion, which starts at zero. Averaged over the set of 21 simulations the growth with time is fitted as $\sqrt{\langle v_r^2\rangle} = 21.5 \,(t/10\,{\rm Gyr})^{1/2}\kms$, for the inner five rings which are most relevant for current observational comparisons. The $\sqrt{t}$ behaviour is expected in a random walk process. The increase in $\langle v_r^2\rangle$ measures radial energy changes, but does not capture the degree to which this energy is coherent long wave motions of the entire ring, or, is effectively thermalized in small scale motions.  For practical observational tests the z-folds of configuration space, Fig.~\ref{fig_xy}, will appear to be a dispersion in velocities in the ring. 

A series of tests finds that the most massive 100 sub-halos, with a minimum mass of $8\times 10^8\msun$, which collectively contain about 62\% of the mass of the 2513 sub-halos create $\sim95$\% of the increase in velocity dispersion of the star streams. It is interesting to note that the gravitational cross-section, $\sim \pi[2GM/V^2]^2$, of the 30 most massive sub-halos is approximately 95\% of the sum of all sub-halos, whereas the volume filling factor defined by the core radii of the top 100 is closer to being the 95\% of the total. The low-mass sub-halos have precisely the same dynamical effects as the larger halos, but on smaller scales so play a reduced role in increasing the velocity dispersion but are important in the small scale scrambling of the streams.

The ring z-folds can be understood as the development from an initial small wave-like distortion induced as a sub-halo crosses the orbit of a stream.  If the initial distortion is a trailing inward perturbation it will rotate faster and gradually overtakes a leading outward perturbation, developing into the z-fold that will be observationally apparent as a stream widening or velocity dispersion increase. An individual fold is readily visible for about a rotation period. Distortions in the opposite sense tend to smooth out with time.

\begin{figure}
\epsscale{1.0}
\plotone{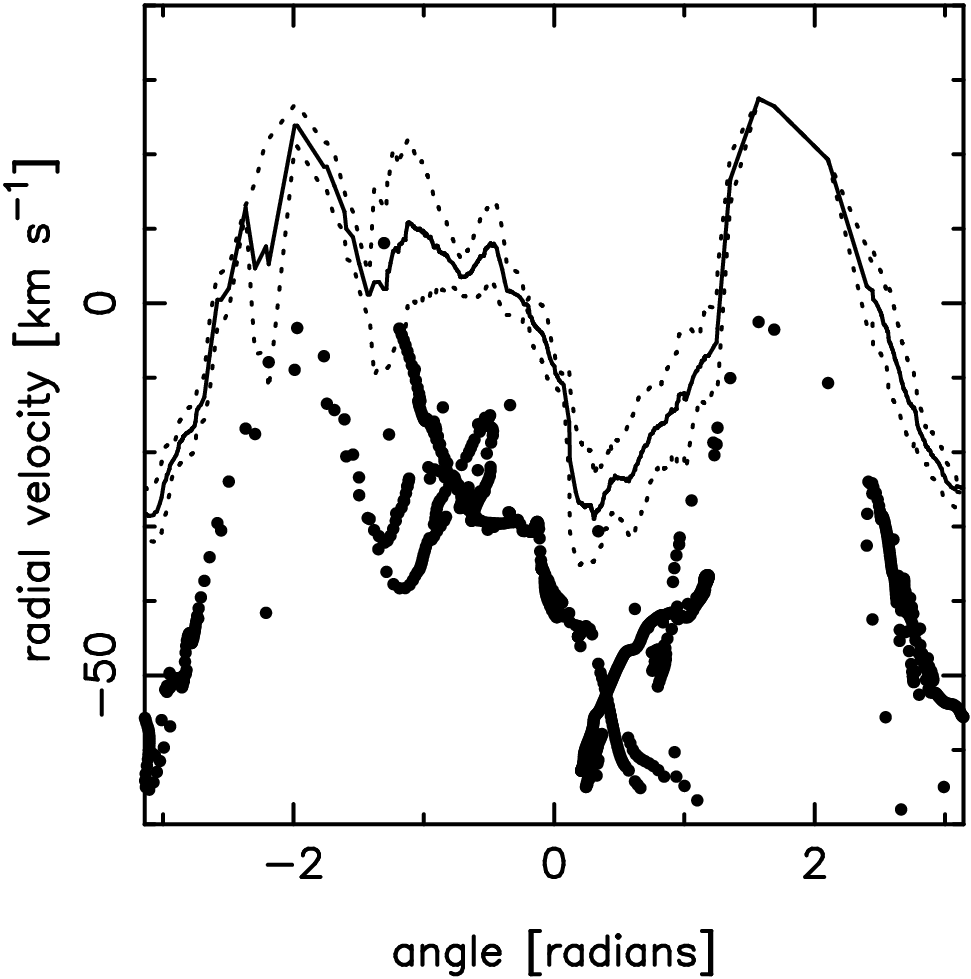}
\caption{The radial velocities around a ring, with angle plotted in radians. The dots show the radial velocities, offset by 30 \kms, of the individual particles in a ring initially at 34 kpc at 8.3 Gyr. The solid line shows the mean velocity averaged over 5.625 degrees. The dotted lines show the velocity dispersion around the circle.  Note that the radial kinetic energy of the entire ring is much larger than this average velocity dispersion around the local mean. The same plot using only 100 sub-halos has the samel large-scale features and but less small scale dispersion.
\label{fig_avr}}
\end{figure}

A practical approach to measuring the local velocity dispersion, often used in observational work, is to compute a local mean velocity by averaging over some range of angles, then calculate the local velocity dispersion from the differences relative to this local mean. For the measures here we use an angular range of $90^\circ$ down to $2^\circ$ for averaging. The mean velocity and local velocity dispersion measure as a function of angle around the circle for a specific realization is shown in Figure~\ref{fig_avr}.

\begin{figure}
\epsscale{1.0}
\plotone{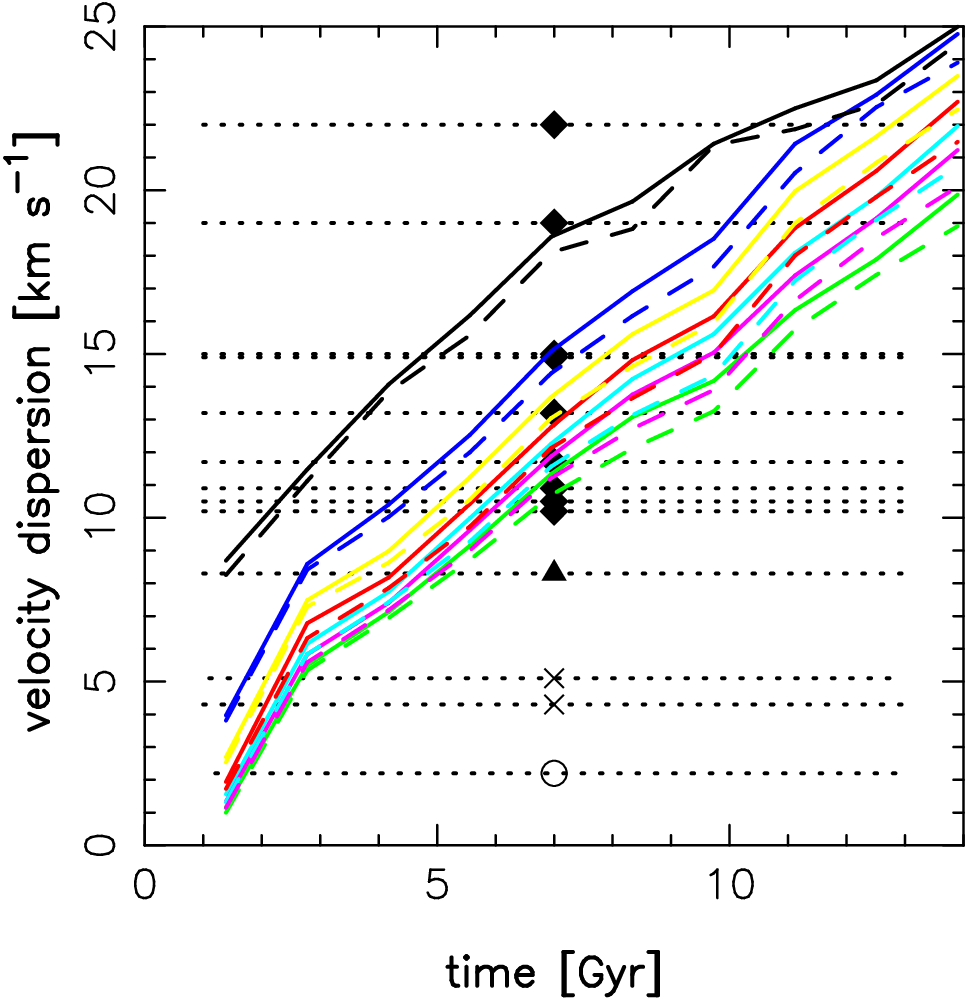}
\caption{The uppermost lines show the rise with time of $\sqrt{v_r^2}$ and the lines below show the RMS velocity dispersion around the local mean stream velocity, averaged over an angular range of 90, 45, 22.5, 11.25, 5.625, and 2 degrees, from top to bottom, respectively. The solid lines are for all 2513 sub-halos and the dashed for the 100 most massive sub-halos. The symbols show star stream velocity dispersions measured in the Milky Way and M31 (circle Pal 5 stream, crosses M31 streams, triangle SGR stream, diamonds ECHOS streams), arbitrarily placed at a dynamical age of 7~Gyr, with dotted lines to indicate that a wide range of ages is possible.
\label{fig_sigrt}}
\end{figure}

The increase with time of the RMS radial velocity dispersion relative to the local mean velocity using the range of averaging angles, ranging from 90 degrees,  or $\pm 45^\circ$, to $2^\circ$, or $\pm 1^\circ$, is shown in Fig.~\ref{fig_sigrt}, again for the inner five rings. The complete set of rings show the same increase but with about a 20\% lower velocity dispersion at any time than the inner five.
The increase with time of the radial velocity dispersion is $\sigma_r= 17.1\, (t/10\,{\rm Gyr})^{0.67}\,\kms$ and to $13.5\, (t/10\,{\rm Gyr})^{0.88}\, \kms$, for averaging angles of 90 and 2 degrees, respectively. In this fit we have excluded data with $t<2$\,Gyr which steepens the slope beyond 1 and does not accurately reflect the majority of the growth rate.

The growth of velocity dispersion with time can be understood as the action of the folding process on an initially cold stream. In the cold stream the increase of velocity dispersion is proportional to a linear rate of interactions. However, as the stream becomes folded the changes begin to resemble a random walk and the growth becomes closer to $\sqrt{t}$. For our 21 simulations there is essentially no correlation between the final overall velocity dispersion and either the total mass of the sub-halos, or, the mass of the single heaviest sub-halo, although the most massive sub-halos do induce the largest scale disturbances. Clearly if we reduced the normalization of the sub-halo $N(M)$ distribution the heating would also be reduced.

\section{DISCUSSION AND CONCLUSIONS}
\nobreak
Galactic star streams are expected to be created with locally small velocity dispersions as a result of their formation from tidally disrupted dwarf galaxies and subsequent cooling of random velocities as the stream elongates and initially nearby stars are separated by their velocity differences leading to reduced local velocity dispersion. Star stream folding sets a minimum velocity as a function of age of the stream, Fig.~\ref{fig_sigrt}, to which details of the disrupted satellite and subsequent orbital evolution will add. Eventually it will be valuable to model the details of individual tidal streams to establish more precise limits. Here we undertake a preliminary assessment of the situation.

Over the past few years a substantial body of kinematic observations of Milky Way and M31 star stream radial velocity dispersions has been developed. \citet{Monaco} report that the velocity dispersion in the trailing tail of the Sagittarius stream are $8.3\pm0.9\kms$, somewhat smaller than the $10.4\pm 1.3\kms$ reported in \citet{Majewski:04}.  The ECHOS group identifies high confidence metal poor streams in the Milky Way with velocity dispersions ranging from 10.5 to 22 \kms\ \citep{Schlaufman:09}. The lowest velocity dispersion reported is an astonishingly low 2.2 \kms\ (or 4.7 if two possible outliers are included) for the tidal stream of Pal 5 \citep{Odenkirchen}.  For two streams in M31, \citet{Chap:08} report velocity dispersions of $5.1\pm2.5$ and $4.3\pm1.6 \kms$.

It is notable that very low velocity dispersion streams exist and that the velocity dispersions of currently measured streams span essentially the same range of velocity dispersions that sub-halo heating creates. On the other hand, if the lowest velocity dispersion streams are older than about 3~Gyr, then LCDM sub-halos would be ruled out. The least certain parameter for the comparison is the dynamical ages of the star streams. Most of the streams are composed of metal poor stars and hence come from old stellar systems, but the stellar age only sets an upper limit on the dynamical age of the stream.  It certainly is remarkable that a few streams have velocity dispersions under 5 \kms, which means that at least some streams are created with sufficiently low internal velocity dispersions that the heating will dominate their eventual evolution. The emerging picture is that some of the current measurements of stream velocity dispersions are only compatible with the velocities that LCDM sub-structured halo allows if the star stream has been orbiting in the galaxy for less than approximately 3 Gyr.  A firm conclusion will require more extensive orbit modelling in well matched simulations. 

Ultimately there is a statistical element to the star stream disruption depending on the details of the sub-halo orbits, so a conclusive study will require modelling a well defined complete sample, ideally for several galaxies. The methods of star stream discovery often use a color-magnitude relation as a key element to pick out prospective stream stars. Future proper motion and radial velocity surveys will allow a completely kinematic selection which will also help with the determination of dynamical ages in detailed orbit modelling.

Star stream kinematics and structure have been shown to be a simple and practical test for dark sub-halos in galactic halos. In the case of a very cold stream from time to time it may be possible to catch it in either position or velocity space at the time at which a ``z-fold" has just developed which provides an additional test. However, the internal velocity dispersion of the stream will blur this out in many cases, but there may be detectable velocity dispersion bulges along the stream.  
More velocity and width measurements of star streams, along with age and orbt modelling will be able to put strong limits on the presence of sub-halos in the halo. The best test will come from the lowest velocity dispersion streams, which likely come from relatively small systems and hence have a relatively low density of stars in the stream. The current observational situation, where several of the streams kinematically studied have impressively low velocity dispersions, indicates a problem with the LCDM predicted numbers of sub-halos if the star streams have dynamical ages of 3~Gyr or older.

\acknowledgements

This research is supported by NSERC and the Canadian Institute for Advanced Research. Comments from the anonymous referee usefully improved this paper.

\end{document}